\documentclass[aps,prl,twocolumn,groupedaddress,showpacs]{revtex4}
\usepackage{graphicx}
\usepackage{amsmath}

\begin{document}

\title{Doping of Bi$_2$Te$_3$ using electron irradiation}

\author{C. W. Rischau$^1$, B. Leridon$^2$, B. Fauqu\'e$^2$, V. Metayer$^1$, and C. J. van der Beek$^1$\footnote{Corresponding author: kees.vanderbeek@polytechnique.edu}}
\affiliation{$^1$Laboratoire des Solides Irradi\'es (CNRS-CEA/DSM/IRAMIS), Ecole Polytechnique, 91128 Palaiseau Cedex, France\\
$^2$Laboratoire de Physique et d'Etude des Mat\'eriaux (UPMC-CNRS), Ecole Sup\'erieure de Physique et de Chimie Industrielles, 10 rue Vauquelin, 75005 Paris Cedex 05, France\\}


% 61.80.-x	Irradiation effects in solids
% 71.18.+y	Fermi surface: calculations and measurements; effective mass, g factor
% 71.70.Ej Spin-orbit coupling, Zeeman and Stark splitting, Jahn-Teller effect

\date{\today}

\begin{abstract}
Electron irradiation is investigated as a way to dope the topological insulator Bi$_2$Te$_3$. For this, p-type Bi$_2$Te$_3$ single crystals have been irradiated with 2.5 MeV electrons at room temperature and electrical measurements have been performed in-situ as well as ex-situ in magnetic fields up to 14 T. The defects created by irradiation act as electron donors allowing the compensation of the initial hole-type conductivity of the material as well as the conversion of the conductivity from p- to n-type. The changes in carrier concentration are investigated using Hall effect and Shubnikov-de Haas (SdH) oscillations, clearly observable in the p-type samples before irradiation, but also after the irradiation-induced conversion of the conductivity to n-type. The SdH patterns observed for the magnetic field along the trigonal axis can be entirely explained assuming the contribution of only one valence and conduction band, respectively, and Zeeman-splitting of the orbital levels. 
\end{abstract}

\maketitle

\section{\label{intro}Introduction}
Bismuth Telluride Bi$_2$Te$_3$ is a narrow-band gap semiconductor (energy gap E$_g$=0.13 eV \cite{Sehr:1962}) which has been extensively studied because of its excellent thermoelectric properties. Recently, it has been discovered that Bi$_2$Te$_3$ as well as Bi$_2$Se$_3$ and Sb$_2$Te$_3$ belong to a group of materials called three-dimensional strong topological insulators (TI) \cite{Moore:2007,Fu:2007a,Fu:2007b}. Surprisingly, these were predicted to be insulating in the bulk, while their surface hosts topologically protected metallic states that can be described as massless Dirac fermions. These conducting surface states have been first observed using surface-sensitive methods like Angle-resolved photoemission spectroscopy (ARPES) \cite{Hsieh:2008} or scanning tunneling microscopy (STM) \cite{Roushan:2009}. However, finding unambiguous evidence of the surface states using transport experiments has proven to be difficult. Namely, in most TI samples charge transport is dominated by a high bulk conductivity due to residual carriers arising from self-doping by point defects in the material, which complicates the identification of the surface contribution to the conductivity \cite{Ren:2010,Qu:2010,Taskin:2011}. So far, successful approaches to suppressing the bulk conductivity include chemical doping \cite{Ren:2010}, electrical gating \cite{Sacepe:2011} or the increase of the surface-to-volume ratio by fabricating thin films or nanowires \cite{Peng:2010}. However, if any future applications of TIs in technology are to be realized, achieving an insulating bulk state in these materials remains a necessary challenge. Here, we investigate electron irradiation-induced defects as a method to reduce the bulk conductivity of the topological insulator Bi$_2$Te$_3$.\\
Bi$_2$Te$_3$ has a tetradymite crystal structure (space group $D^{5}_{3d}$ ($R\bar{3}m$)) with a rhombohedral primitive unit cell which contains five atoms in total, two Bi atoms on equivalent sites and three Te atoms which are distributed on two non-equivalent sites denoted Te1 and Te2. Bi$_2$Te$_3$ has a layered structure and can be described as a stack of so-called \textit{quintuple layers} along the trigonal axis, which each contain five atomic layers in a Te1-Bi-Te2-Bi-Te1 pattern. The bonding between the Te1 layers belonging to two adjacent quintuple layers is of van der Waals type, whereas the bonding within the quintuple layers is covalent. The carrier type in Bi$_2$Te$_3$ depends strongly on the intrinsic point defects present in the material. It is generally accepted that the most prominent point defects in as-grown Bi$_2$Te$_3$ are the Bi$_\textnormal{Te1}$, Bi$_\textnormal{Te2}$ and Te$_\textnormal{Bi}$ antisite defects, with Bi$_\textnormal{Te1}$ and Bi$_\textnormal{Te2}$ acting as acceptors, whereas Te$_\textnormal{Bi}$ acts as electron donor \cite{Hashibon:2011}. Te vacancies were predicted to act as double donors whereas Bi vacancies act as triple acceptors \cite{Pecheur:1994}. However, due to a higher formation energy of vacancies compared to Bi and Te antisite defects during growth of Bi$_2$Te$_3$ crystals, the doping during growth has been predicted to be almost entirely determined by the antisite defects \cite{Hashibon:2011}.\\
Both the top of the highest valence band and the bottom of the lowest conduction band of Bi$_2$Te$_3$ are generally described by a six-valley model with non-parabolic hole and electron pockets, respectively, which are located pairwise in the three mirror planes of the Brillouin zone containing the bisectrix and trigonal axis \cite{Drabble-58a,Drabble-58b}. Thus, the effective mass tensor and the electric transport properties are expected to be highly anisotropic. K\"ohler determined the inverse effective mass tensor $\boldsymbol{\alpha}=\alpha_{ij}/m_e$ ($m_e$: free electron mass) from Shubnikov-de Haas (SdH) oscillations applying an ellipsoidal 6-valley model for the valence band edge ($\alpha_{11}=32.5$, $\alpha_{22}=4.81$, $\alpha_{33}=9.02$ and $\alpha_{23}=4.15$ \cite{Koehler:1976}) and the conduction band edge ($\alpha_{11}=46.9$, $\alpha_{22}=5.92$, $\alpha_{33}=9.5$ and $\alpha_{23}=4.22$ \cite{Koehler:1976a}) with axes 1, 2 and 3 refering to the binary \textbf{n}, bisectrix \textbf{s} and trigonal axis \textbf{c}, respectively. Moreover, from the observation of a second SdH oscillation period for \textbf{B}$\parallel$\textbf{c} on p-type samples with hole densities higher than 4 x 10$^{18}$ cm$^{-3}$, the filling of a second six-valley valence band, located 15 - 20 meV below the highest valence band, has been concluded for this range of hole densities \cite{vonMiddendorff:1972,Sologub:1975,Koehler:1976}.\\
Previous irradiation studies on Bi$_2$Te$_3$ have been carried out using 7.5 MeV protons at room temperature \cite{Chaudhari:1966,Chaudhari:1967}, or 5 MeV electrons at 250 K \cite{Karkin:1998}. Hall effect measurements on the irradiated samples have shown that it is in fact possible to convert the conduction of initially p-type Bi$_{2}$Te$_3$ to n-type using irradiation induced defects. From proton irradiations of Bi$_2$Te$_3$, Chaudhari and Bever estimated the threshold displacement energy $E_d$ for a lattice atom to be in the range of 7.5 - 12.5 eV for Te1 atoms and 15 - 25 eV for Te2 and Bi atoms \cite{Chaudhari:1966}. It was argued that the displacement threshold for Te1 atoms is lower than for Te2 or Bi atoms because of the weaker van der Waals bonding between Te1 layers compared to the stronger covalent bonding within the quintuple layers. Annealing at 350 K of samples displaying n-type conduction after irradiation was found to restore the initial p-type conduction \cite{Karkin:1998}.\\
We have irradiated p-type Bi$_{2}$Te$_3$ single crystals at room temperature with 2.5 MeV electrons, and studied the compensation of the acceptor-type defects by monitoring the resistivity of the samples in-situ during irradiation. Changes in carrier density were deduced from ex-situ magnetoresistance measurements in fields up to 14 T, by analyzing the Hall effect and SdH oscillations, detected for all irradiation doses in both the p- and n-type regime. This, together with the excellent agreement of the extracted carrier effective masses and Fermi energies with those measured on Bi$_2$Te$_3$ doped by chemical means \cite{Koehler:1976,Koehler:1976a}, attests to the fact that doping is the main effect of the irradiation, and that charge carrier scattering is secondary. Moreover, the analysis of the SdH effect allows a coherent interpretation of the Fermi surface in p- and n-type Bi$_2$Te$_3$ in terms of a six-valley model and Zeeman splitting, without the need to invoke the contribution of either a second valence or conduction band or surface carriers.

\section{\label{exp}Methods}
P-type Bi$_{2}$Te$_3$ single crystals (see \cite{Hruban:2011} for fabrication details) were cut into bars with lateral dimensions of 1 x 3 mm$^2$ using a wire saw and then thinned down to a thickness of 10 to 40 $\mu$m using scotch tape. Gold contact pads were evaporated on freshly cleaved surfaces, and contacted using silver paint and Au wires. All samples were irradiated at room temperature with 2.5 MeV electrons at the \textit{SIRIUS} Pelletron accelerator facility of the \textit{Laboratoire des solides irradi\'es}. The four-probe resistance of the samples was measured in-situ during irradiation using an AC-technique. After irradiation the samples were stored in liquid nitrogen to prevent any long-term annealing. They were then measured ex-situ between 1.9 and 300 K using a Quantum Design 14 T Physical Properties Measurement System (PPMS). All measurements were performed with the magnetic field oriented along \textbf{c}. Prior to irradiation, all samples showed a similar temperature and field dependence of resistivity and Hall coefficient, differences in the absolute values being do to the uncertainty in the determination of the contact distances. In order to compare all samples, their initial resistivity has been normalized to the average resistivity of all unirradiated samples.

\section{\label{results}Results \& Discussion}
\subsection{Resistivity and low-field Hall effect}
\begin{figure*}
	\centering
		\includegraphics[width=0.92\textwidth]{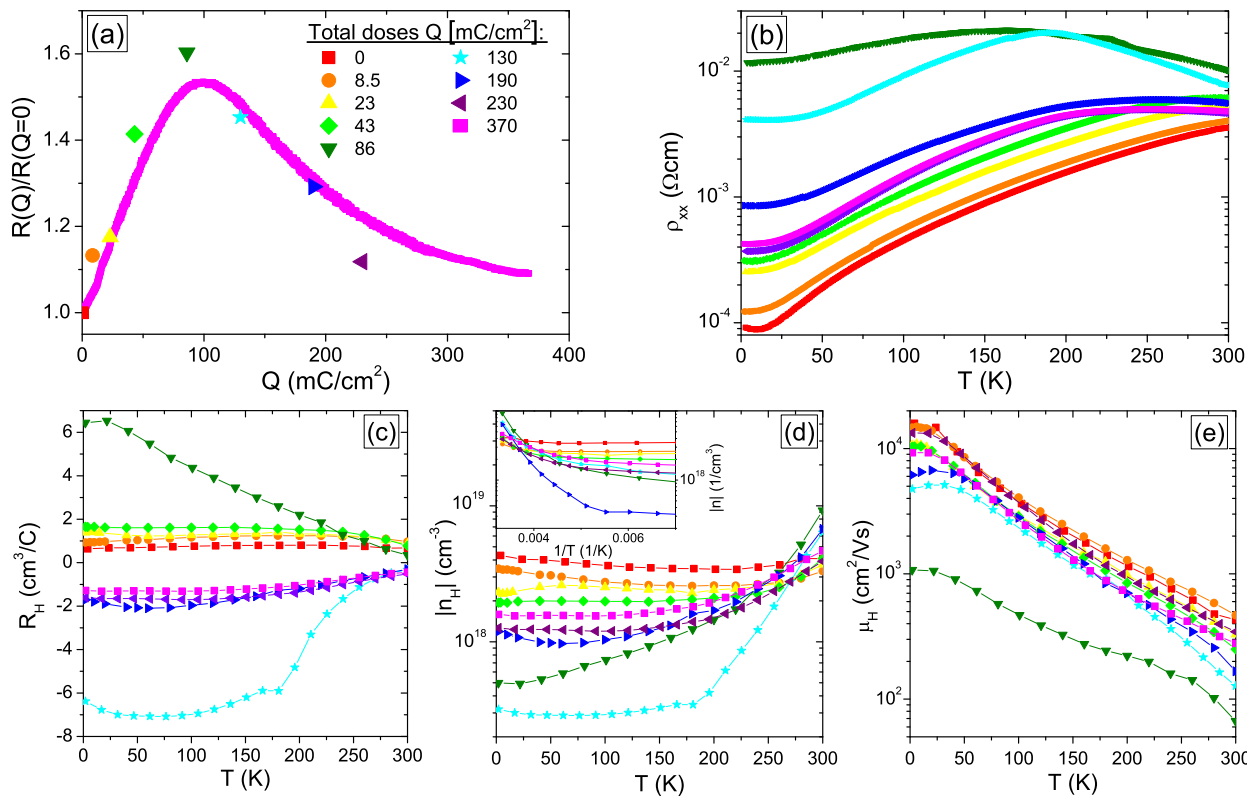}
	\caption{\textbf{(a)} Normalized resistance $R(Q)/R(Q=0)$ of a Bi$_2$Te$_3$ single crystal, measured in-situ during 2.5 MeV electron irradiation at room temperature, as a function of electron dose $Q$ (magenta squares). The different other symbols represent the $R(Q)/R(Q=0)$ values obtained for different samples, irradiated to different total doses Q. \textbf{(b)} Resistivity $\rho_{xx}$ and \textbf{(c)} low-field Hall coefficient $R_{H}$ of the same samples measured ex-situ as a function of temperature $T$. \textbf{(d)} Absolute values of the Hall carrier concentration $\left|n_{H}\right|$ and \textbf{(e)} the Hall mobility $\mu_{H}$, both calculated assuming only one carrier type. The inset of (d) shows $\left|n_{H}\right|$ as a function of $1/T$.}
	\label{fig:figure1}
\end{figure*}

\begin{figure}
	\centering
		\includegraphics[width=0.47\textwidth]{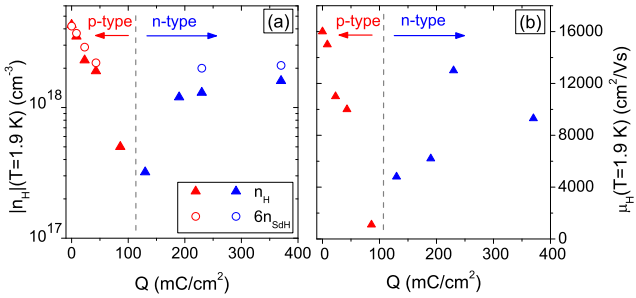}
	\caption{\textbf{(a)} Absolute value of the carrier concentrations $n_H$ and $6n_{SdH}$ obtained from low-field Hall effect and SdH oscillations at 1.9 K, respectively, as a function of dose $Q$. \textbf{(b)} Hall mobility $\mu_H$ at 1.9 K as a function of $Q$.}
	\label{fig:figure2}
\end{figure}

Figure \ref{fig:figure1} (a) shows the resistance of a Bi$_2$Te$_3$ single crystal measured in-situ, normalized to the initial resistance measured before irradiation, as a function of electron dose $Q$. The resistance first increases linearly up to a dose of around 50 mC/cm$^2$ until it reaches a maximum of 1.5 - 1.6 times its initial value at $Q \approx$ 80 - 90 mC/cm$^2$. With further irradition, the resistance decreases again, nearly reaching its initial value at a dose of 370 mC/cm$^2$. A similar behaviour was observed for several samples irradiated to different total doses $Q$. For these, only the last points of the measured $R(Q)/R(Q=0)$ curves are shown in Fig. \ref{fig:figure1} (a).\\
Figure \ref{fig:figure1} (b) shows the resistivity of the irradiated samples as well as that of an unirradiated sample as a function of temperature. The resistivity of the virgin sample shows the metallic temperature dependence expected for p-type Bi$_2$Te$_{3}$, i.e., $\rho_{xx}$ decreases with decreasing temperature. For the samples irradiated to total doses of 8.5, 23 and 43 mC/cm$^{2}$, $\rho_{xx}$ still displays a metallic temperature dependence, even if the absolute values of the resistivity are increased over the entire temperature range. After irradiation with 86 mC/cm$^{2}$ the material is near optimum compensation and $\rho_{xx}$ shows a more complicated temperature dependence. It first increases with decreasing $T$, reaches a maximum at around 160 K, before decreasing to a resistivity of 12 m$\Omega$cm at 1.9 K which corresponds to an increase by approximately two orders of magnitude compared to the virgin sample. It should be noted that this value of $\rho_{xx}$(1.9 K) is comparable to that obtained on non-metallic Bi$_2$Te$_3$ samples (12 m$\Omega$cm) cut from crystals grown with a weak compositional gradient \cite{Qu:2010} or lightly doped Bi$_{2-x}$Tl$_x$Te$_3$ samples (28 m$\Omega$cm for $x=0.1$) \cite{Chi-13}. However, it is much lower than that obtained for heavy chemical doping in Bi$_2$Te$_{3-x}$Se$_{x}$ (1 $\Omega$cm for $x=0.9-1$) \cite{Ren:2010,Akrap:2013}. After further irradiation to 130 mC/cm$^2$, $\rho_{xx}$ shows a similar temperature dependence, but compared to the sample irradiated with 86 mC/cm$^{2}$, the absolute values of the resistivity have decreased again over the entire investigated temperature region. For samples irradiated to doses higher than 190 mC/cm$^{2}$, $\rho_{xx}$ shows again a metallic temperature dependence below 260 K.\\
Fig. \ref{fig:figure1} (c) shows the low-field Hall coefficient $R_{H}$ as a function of temperature, determined from the slope $\frac{\Delta\rho_{yx}}{\Delta B}$ of the Hall resistivity $\rho_{yx}$ around zero field ($-0.2$ T $\lesssim B \lesssim 0.2$ T) using a linear fit. The virgin sample displays a positive and almost temperature independent $R_{H}(T)$. After irradiation up to $Q=86$ mC/cm$^{2}$, $R_{H}$ is still positive over the entire temperature range, but now increases linearly with decreasing T up to a value of 6.5 cm$^3$/C at 1.9 K, even if the value at room temperature remains nearly the same as for the virgin sample. After irradiation with 130 mC/cm$^{2}$, $R_{H}$ has changed sign, i.e., the conduction has changed from p- to n-type. With further irradiation $R_{H}$ increases, but remains negative.\\
In case of an anisotropic Fermi surface, the Hall carrier density $n_H$ is given by the low-field Hall coefficient $R_{H}$ via $n_{H}=ra/(eR_{H})$ with the Hall scattering factor $r$ and an anisotropy factor $a$. The factor $ra$ has been estimated by Drabble from galvanomagnetic measurements as $ra=0.514$ and 0.32 for p- and n-type Bi$_2$Te$_3$, respectively \cite{Drabble-58b,Drabble-58a}. It should be noted, that since $ra{\rightarrow}1$ for high magnetic fields, one can alternatively use the high-field saturation value $R_{H}^{\infty}$ of the Hall coefficient (if determinable) to calculate the carrier density $n_{H}^{\infty}=1/(eR_{H}^{\infty})$, as has been done in previous works on Bi$_2$Te$_3$ \cite{Koehler:1976,Koehler:1976a,vonMiddendorff:1972,Sologub:1975,Koehler:1976c,Koehler:1976b}. Comparison of the carrier densities calculated from the low- and high-field Hall coefficient given in table \ref{table1} confirms that both are comparable. Figs. \ref{fig:figure1} (d) and (e) show the Hall carrier density $|n_H|$ and the Hall mobility $\mu_H=1/(en_{H}\rho_{xx})$, respectively, as a function of temperature. The virgin sample shows a nearly temperature-independent $n_H$. After irradiation with 86 mC/cm$^{2}$, $n_H(T=1.9$ K) decreased to 5.0 x 10$^{17}$ cm$^{-3}$, i.e., a decrease of one order of magnitude compared to the value in the unirradiated sample, 4.3 x 10$^{18}$ cm$^{-3}$. However, it should be noted that $n_{H}=ra/(eR_{H})$ does not give the exact carrier concentration for samples near optimum compensation, since two types of carriers with comparable concentrations are present in this region. The inset in Fig. \ref{fig:figure1} (d) diplays $|n_H|$ as a function of 1/T for high temperatures. For temperatures above 280 - 300 K, $n_H$ seems to show an activated behaviour which is the most pronounced for samples irradiated to 86 and 130 mC/cm$^{2}$, i.e. the region of activated behaviour extends to lower T. However, an upper temperature limit of 300 K for the measurements is not high enough in order to reliably determine values for the activation energies. Since an annealing of electron-irradiated defects at 350 K has been reported \cite{Karkin:1998}, we have not measured our samples at temperatures higher than 300 K. Figs. \ref{fig:figure2} (a) and (b) show $n_H$ and $\mu_H$ at 1.9 K as a function of Q. Both, $n_H$ and $\mu_H$ decrease as one approaches the dose of optimal compensation. Interestingly, $\mu_H$ increases again once the conduction changed to n-type. A similar dependence of the mobility as a function of dose, i.e. carrier concentration, was observed during irradiation of PbSnTe \cite{Green:1976a} and Hg$_{1-x}$Cd$_x$Te \cite{Green:1976b} and was explained with a model based on ionized impurity scattering.\\

\subsection{High-field magnetoresistance}
\begin{figure}
	\centering
		\includegraphics[width=0.47\textwidth]{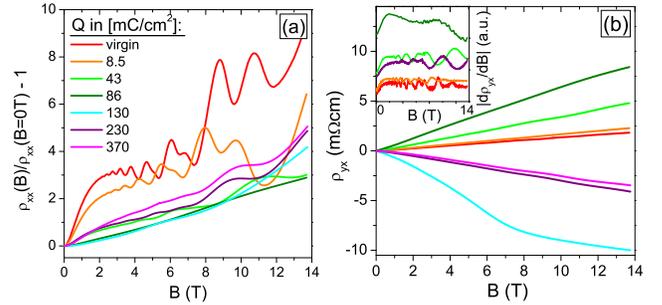}
	\caption{\textbf{(a)} Normalized magnetoresistance $\rho_{xx}(B)/\rho_{xx}(B=0$ T$) - 1$ measured at 1.9 K as a function of magnetic field $B$ after different total irradiation doses $Q$. \textbf{(b)} Hall resistivity $\rho_{yx}$ at 1.9 K as a function of $B$ with the inset showing $\left|d\rho_{yx}/dB\right|$ vs. $B$.}
	\label{fig:figure3}
\end{figure}

Figure \ref{fig:figure3} (a) depicts the normalized magnetoresistance \mbox{$\rho_{xx}(B)/\rho_{xx}(B=0$ T$)-1$} as function of magnetic field $B$ for the virgin and selected irradiated samples at 1.9 K. The normalized magnetoresistance of the unirradiated sample shows a non-linear background with a superimposed pattern of Shubnikov-de Haas (SdH) oscillations. For irradiation up to 86 mC/cm$^2$, the normalized magnetoresistance decreases as a function of electron dose and the background becomes more linear. The oscillations become less pronounced, yet changes regarding the positions and the period of the oscillations are clearly visible. After irradiation doses of 130 mC/cm$^2$ and higher, the normalized magnetoresistance increases again and still shows an almost linear $B$ dependence. Furthermore, the SdH oscillations seem to reappear, i.e., become more pronounced again.\\
The Hall resistivity $\rho_{yx}$ as a function of $B$, depicted in Fig. \ref{fig:figure3} (b), shows a linear background at magnetic fields $B \gtrsim$ 1 T on which SdH oscillations are superposed. However, for small magnetic fields up to B $\lesssim$ 1 T, $\rho_{yx}$ has a parabolic field dependence, which can be seen in the plot of $\left|d\rho_{yx}/dB\right|$ vs. B (see inset of Fig. \ref{fig:figure3} (b)). This non-linearity of $\rho_{yx}(B)$ has been previously reported for p-type Bi$_2$Te$_3$, however, only for hole densities higher than 4 x 10$^{18}$ cm$^{-3}$, and been attributed to the presence of a second valence band for these carrier densities \cite{Koehler:1976,Karkin:1998}. It should be noted that we observe this non-linearity for all n- and p-type samples independent of irradiation dose, i.e., for hole densities in the range of $n \approx$ 0.5 - 4.3 x 10$^{18}$ cm$^{-3}$. Therefore, we propose that the observed nonlinearity is not due to a different type of holes originating from a second valence band, but the strong anisotropy of the Fermi surface.

\subsection{Shubnikov-de Haas oscillations}
\begin{figure*}
	\centering
		\includegraphics[width=0.9\textwidth]{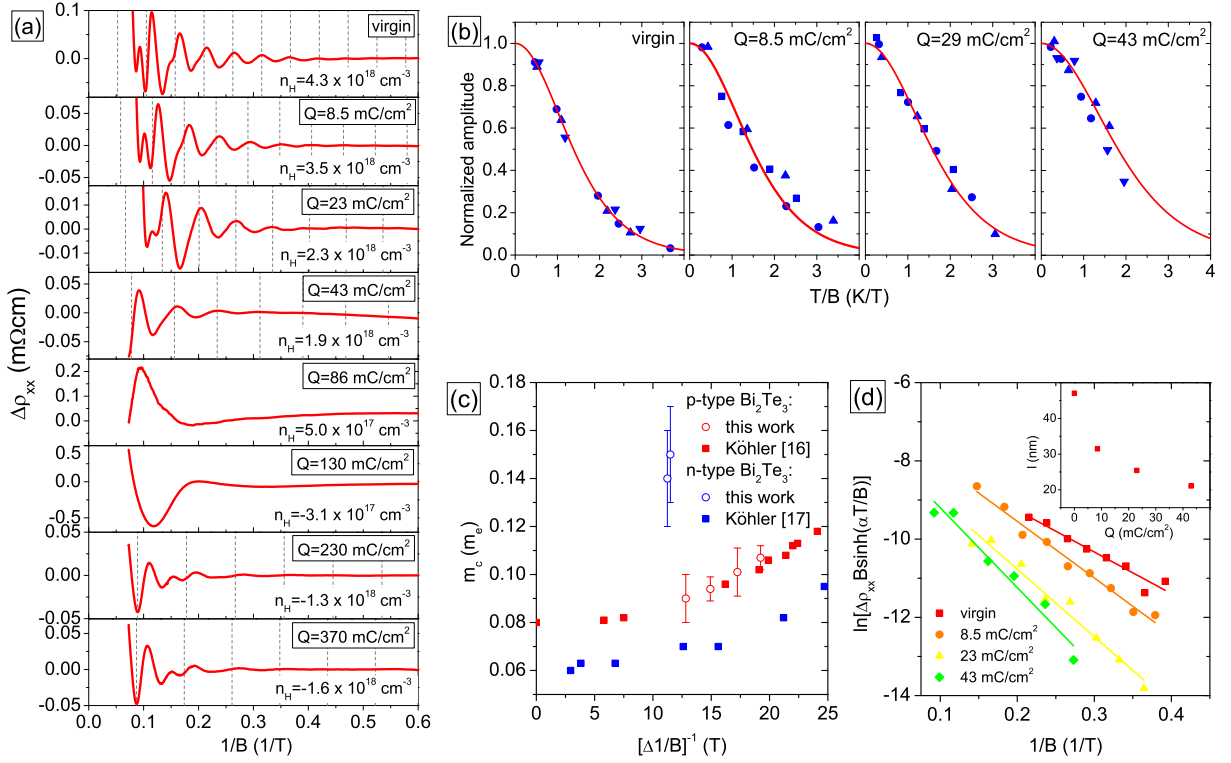}
	\caption{\textbf{(a)} Oscillating part of the resistivity $\Delta\rho_{xx}$ at $1.9$ K for $\textbf{B}\parallel$\textbf{c} as a function of inverse magnetic field $1/B$ for different irradiation doses. For each dose, the vertical gray grid lines indicate the multiples of the oscillation periods $\Delta(1/B)$ given in table \ref{table1}. \textbf{(b)} Temperature dependence of the oscillation amplitude with solid lines representing the fits obtained using Eq. \ref{eq:SdH2}. Different symbols correspond to the analysis of different LLs, i.e., peaks at different $1/B$ positions. In order to compare the temperature dependence of different peaks, the oscillation amplitudes have been normalized to the value of the fit for $1/B{\rightarrow}0$. \textbf{(c)} Cyclotron masses $m_c$ as function of the inverse oscillation period $[\Delta(1/B)]^{-1}$. \textbf{(d)} Dingle plots at $T=1.9$ K and inset showing the mean free path $l$ as a function of dose $Q$.}
	\label{fig:figure4}
\end{figure*}
Fig. \ref{fig:figure4} (a) shows the resistivity $\rho_{xx}$ at 1.9 K after subtraction of a smooth background as a function of $1/B$ for different irradiation doses. The unirradiated p-type sample shows very pronounced SdH oscillations with a double-peak structure at low $1/B$ values being either due to Zeemann-splitting or the superposition of a second oscillation frequency. Assuming a six-valley structure for the highest valence band \cite{Koehler:1976}, only one oscillation frequency is expected for \textbf{B}$\parallel$\textbf{c} whereas for \textbf{B}$\parallel$\textbf{s} or \textbf{B}$\parallel$\textbf{n} two different oscillation frequencies should arise. We propose that the observed splitting for \textbf{B}$\parallel$\textbf{c} on our samples (hole densities $<$ 4 x 10$^{18}$ cm$^{-3}$) is due to Zeeman-splitting. It should be noted, that the appearance of a second oscillation frequency for \textbf{B}$\parallel$\textbf{c} has been previously observed for hole densities $>$ 4 x 10$^{18}$ cm$^{-3}$ and ascribed to the contribution of a second six-valley valence band \cite{Sologub:1975,Koehler:1976}. Assuming Zeeman-splitting, the periodicity of the SdH pattern can be well described with a single oscillation period of \mbox{$\Delta(1/B)=0.052$ 1/T}, with the multiples of $\Delta(1/B)$ (see vertical grid lines in Fig. \ref{fig:figure4} (a)) exactly centered between split peaks.\\
With increasing dose, i.e., decreasing carrier density, the peaks corresponding to higher Landau levels (LLs) become less visible due to both the decrease of the Fermi energy and the increase of irradiation-induced lattice-disorder. For the p-type samples up to total doses of 43 mC/cm$^2$, the periodicity of the SdH pattern can always be described by one single oscillation period $\Delta(1/B)$ that increases with irradiation dose, i.e. decreasing hole concentration (see Tab. \ref{table1}). Even if the peak splitting at lower $1/B$ values becomes less visible with increasing dose, it can be observed for all p-type samples up to 43 mC/cm$^2$. For the samples irradiated to 86 and 130 mC/cm$^2$ one observes just one oscillation maximum and minimum, respectively, which makes analysis impossible. It should be noted that, the mobility estimated from the low-field Hall effect $\mu_H$(1.9 K)=1100 cm$^2$/Vs (see Fig. \ref{fig:figure2}) is still quite high for the sample near optimum compensation ($Q=86$ mC/cm$^{2}$). Interestingly, after further irradiation to 230 and 370 mC/cm$^2$, i.e., increasing electron carrier concentration, a larger number of maxima reappears. For the n-type samples, there seems to be either spin-splitting, as in the case of the p-type samples, or, alternatively, the appearance of a second oscillation frequency due to the contribution of a second type of carrier.\\
\squeezetable
\begin{table*}
\caption{\label{table1}Overview of the carrier concentrations $n_H$ and $n_{H}^{\infty}$ obtained from the low- and high-field Hall coefficient at 1.9 K as well as the parameters obtained from temperature and field dependence of the oscillation amplitude measured at 1.9 K for \textbf{B}$\parallel$\textbf{c} (n.d.: parameter non-determinable due to a too limited number of oscillation maxima). The $m_z$ values were not determined experimentally, but were calculated from the effective mass tensors given in \cite{Koehler:1976} and \cite{Koehler:1976a} for p- and n-type Bi$_2$Te$_3$, respectively. The values of $6n_{SdH}$, the spin-splitting factor $M$ and the $g$-factor were determined by simulations based on Eq. \ref{eq:SdH3}.}
\begin{ruledtabular}
\begin{tabular}{cccccccccccc}
 & \multicolumn{2}{c}{\textit{Hall effect}} & \multicolumn{5}{c}{\textit{Temperature dependence and Dingle analysis}} & \multicolumn{4}{c}{\textit{Fit}} \\
Q 			& $n_{H}$	& $n_{H}^{\infty}$	& $\Delta (1/B)$ 	& $m_{c}$ 	& $\epsilon_F$ 	& $\tau_D$ 			& $T_D$ &  $m_z$ & $6n_{SdH}$ & M & g\\
{}[mC/cm$^2$] & \multicolumn{2}{c}{$[10^{18}$ cm$^{-3}]$}   &  [1/T] 					& [$m_{e}$]  			& [meV]  				& [$10^{-13}$s] & [K] 	&  [$m_{e}$] & $[10^{18}$ cm$^{-3}]$ & & \\
\cline{1-1} \cline{2-3} \cline{4-8} \cline{9-12}
 0 	& 4.3 & 4.6 & 0.052 $\pm$ 0.0005	& 0.107 $\pm$ 0.01 	& 21 $\pm$ 1 & 1.8 $\pm$ 0.2 & 6.5 $\pm$ 0.4 &  0.244 & 4.2 & 0.67,1.33 & 12.5,24.9 \\
 8.5 & 3.5 & 3.8 &  0.058 $\pm$ 0.0005	& 0.101 $\pm$ 0.01 	& 20 $\pm$ 2 & 1.2 $\pm$ 0.1 & 10 $\pm$ 0.5  & 0.228 & 3.7 & 0.67,1.33  & 13.3,26.3\\
 23 & 2.3	& 2.8 & 0.067 $\pm$ 0.001		& 0.094 $\pm$ 0.005 & 18 $\pm$ 1 & 0.97 $\pm$ 0.05 & 12.5 $\pm$ 0.7 & 0.214 & 2.9 & 0.67,1.33 & 14.3,28.3 \\
 43 & 1.9 & 1.7	& 0.078	$\pm$ 0.002		& 0.09 $\pm$ 0.01 	& 16 $\pm$ 2 & 0.72 $\pm$ 0.1 & 15 $\pm$ 2 & 0.178 & 2.2 & 0.69,1.31 & 15.3,29.1 \\
 86 & 0.5 & 0.9 &  $n.d.$	& $n.d.$	& $n.d.$ & $n.d.$ & $n.d.$  & - & - & - & - \\
 130 & -0.31 & $n.d.$ &  $n.d.$	& $n.d.$	& $n.d.$ & $n.d.$ & $n.d.$  & - & - & - & - \\
 190 & -1.2 & -2.1 & $n.d.$	& $n.d.$	& $n.d.$ & $n.d.$ & $n.d.$  & - & - & - & - \\
 230 & -1.3 & -2.1 & 0.089 $\pm$ 0.002	& $n.d.$	& $n.d.$ & $n.d.$ & $n.d.$  & 0.167 & 2.0 & 0.57,1.43 & 17.5,44 \\
 370 & -1.6 & -2.4 &  0.087 $\pm$ 0.002	& $n.d.$	& $n.d.$ & $n.d.$ & $n.d.$  & 0.167 & 2.1 & 0.57,1.43 & 17.5,44 \\
\end{tabular}
\end{ruledtabular}
\end{table*}
The SdH period $\Delta (1/B)$ can be expressed as
\begin{equation}
\Delta\left(\frac{1}{B}\right) = \frac{2\pi e}{\hbar S_F}=\frac{e\hbar}{m_{c}\epsilon_F}\\
\label{eq:SdH1}
\end{equation}
with $S_F$ the cross section of the Fermi surface perpendicular to the direction of the field $\textbf{B}$, $m_{c}$ the cyclotron mass, $\hbar$ the Planck constant and $\epsilon_F$ the Fermi energy (measured from the band edge). Using the standard Lifshitz-Kosevich theory \cite{Shoenberg:1984}, the temperature dependence of the oscillation amplitude $\Delta\rho_{xx}$ of a fixed peak can been fitted as a function of $T/B$ using
\begin{equation}
\Delta\rho_{xx}(T/B)=\alpha T/(B \sinh(\alpha T/B))
\label{eq:SdH2}
\end{equation}
with $\alpha=2\pi^2m_{c}k_{b}/(\hbar e)$, allowing for the extraction of $m_c$. Fig. \ref{fig:figure4} (b) shows the temperature dependence of the oscillation amplitude as well as the corresponding fits to Eq. \ref{eq:SdH2} for the p-type samples for which this analysis was possible. Similar to K\"ohler's work, cyclotron masses determined from visibly split peaks showed a strong dependence on the $1/B$ position of the analysed peak; therefore, only non-split peaks were used to determine $m_c$. The plot of $m_c$ as a function of the inverse oscillation period $[\Delta(1/B)]^{-1}$ in Fig. \ref{fig:figure4} (c) shows an decrease of $m_c$ with decreasing $[\Delta(1/B)]^{-1}$, i.e., increasing dose and decreasing hole concentration, respectively. This confirms the non-parabolic energy dispersion $\epsilon(\textbf{k})$ of the highest valence band in p-type Bi$_2$Te$_3$. Furthermore, Fig. \ref{fig:figure4} (c) depicts the cyclotron masses found by K\"ohler on a series of p- and n-type samples with doping realized during crystal growth \cite{Koehler:1976,Koehler:1976a} showing a good agreement with the values found on our irradiated p-type samples. However, there is a strong deviation of the $m_c$ values extracted from samples irradiated to 230 and 370 mC/cm$^2$. This probably stems from the error introduced by spin-splitting of the analysed peaks. Unlike the work of K\"ohler, the higher non-split LLs are not visible.\\
The field dependence of the oscillation amplitude can be used to determine the Dingle scattering time $\tau_D$ by plotting $\ln \left[ \Delta \rho_{xx} B \sinh(\alpha T/B)\right]\propto \pi m_c/(e\tau_{D}B)$ as a function of $1/B$ for a fixed $T$ (\textit{Dingle plot}) \cite{Shoenberg:1984}. Fig. \ref{fig:figure4} (d) shows the Dingle plots for the unirradiated and some of the irradiated p-type samples at $T = 1.9$ K. For all other irradiated samples, the number of observed peaks was not sufficient to carry out this analysis. The scattering time $\tau_D$ corresponds to a Dingle temperature $T_D=\hbar/(2 \pi k_{B}\tau_{D})$ and can be used to estimate the mean free path $l=v_F\tau_D$ with $v_F$ the Fermi velocity calculated from Eq. (\ref{eq:SdH1}) and $\epsilon_F=\frac{1}{2}m_c v_{F}^{2}$. The mean free path as a function of $Q$ is shown in the inset of Fig. \ref{fig:figure4} (d) and seems to saturate at about half of its initial value at $Q=43$ mC/cm$^2$. It should be noted, that up to this dose, $n_H$ also decreased by a factor two, indicating that the effects of disorder and the decrease of the Fermi energy are comparable in this dose range. All parameters obtained from the analysis of the temperature and field dependence of the oscillation amplitude are given in table \ref{table1}.\\
\begin{figure*}
	\centering
		\includegraphics[width=0.99\textwidth]{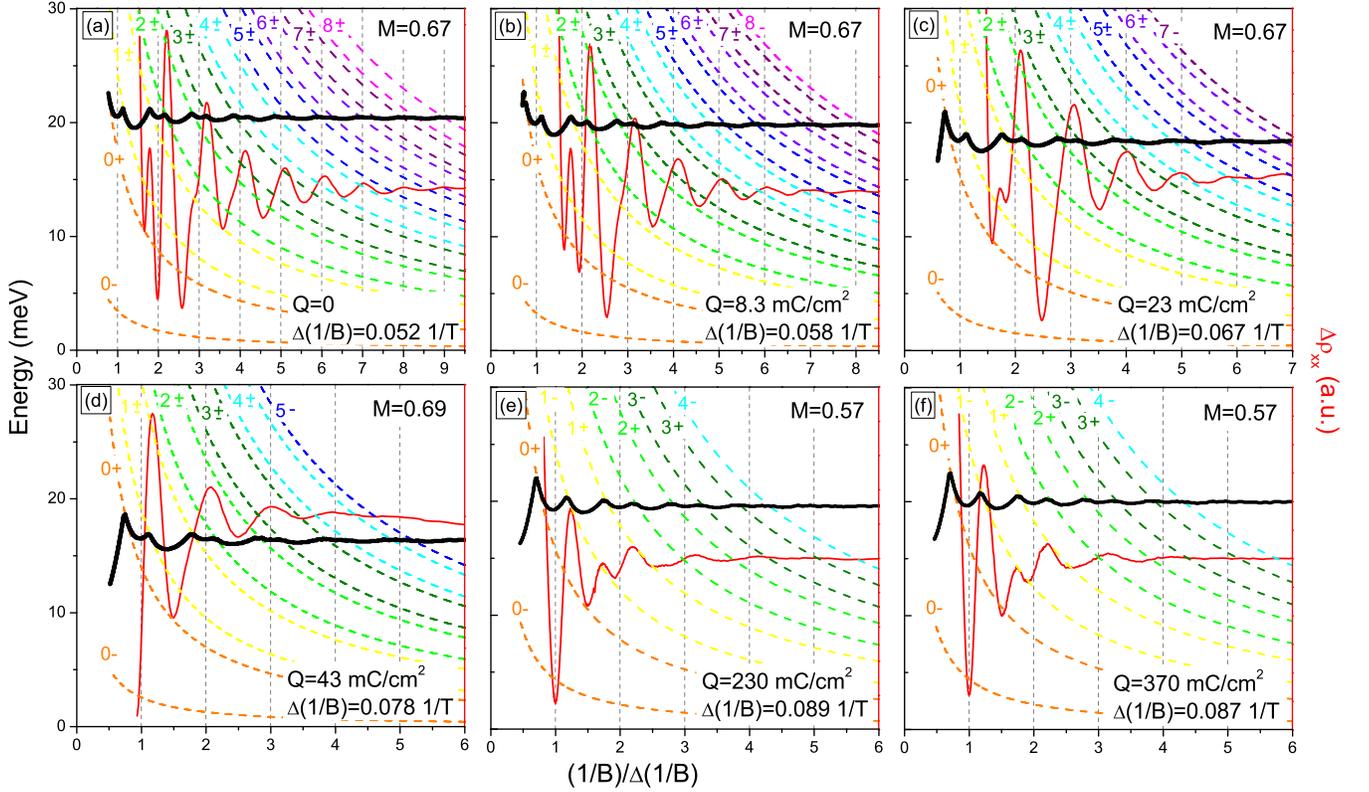}
	\caption{Fermi energy $\epsilon_F$ calculated with Eq. \ref{eq:SdH3} (black line, left scale) and energy of the split Landau levels (colored lines with LL indices, left scale) as well as the oscillating part of the measured resistivity $\Delta\rho_{xx}$ (red line, right scale) as a function of $(1/B)/\Delta(1/B)$ for the virgin sample, as well as samples irradiated to different doses $Q$. The corresponding fitting parameters are given in Tab. \ref{table1}.}
	\label{fig:figure5}
\end{figure*}
Since the Fermi surfaces of p- and n-type Bi$_2$Te$_3$ are non-spherical and non-parabolic, an estimation of the carrier density $n_{SdH}$ from the SdH oscillations is not obvious. In order to calculate $n_{SdH}$ as well as to investigate the origin of the double-peak structure observed at high $B$, we use the following model including spin-splitting to simulate the LL spectra (for details see Appendix). In this model, the carrier density $n_{SdH}$ of one single ellipsoidal hole- or electron-pocket is given by
\begin{equation}
n_{Sdh} = \frac{1}{2\pi^2 l_{B}^{2}} \sum_{i=0}^{i_{max}}\sum_{\pm M}\sqrt{\frac{2m_z}{\hbar^2}\left[\epsilon_F-(i+ \frac{1}{2} \pm \frac{M}{2})\hbar\omega_c )\right]}
\label{eq:SdH3}
\end{equation}
with $l_B=\sqrt{\hbar/eB}$ and $m_z$ the longitudinal mass. The spin-splitting factor $M=\frac{E_Z}{\hbar \omega_c}=g\frac{m_c}{2m_0}$ describes the ratio of the Zeeman energy $E_Z$ to the cyclotron energy $\hbar\omega_c$ and is related to the g-factor $g$. The conductivity $\sigma_{xx}$ is enhanced each time the Fermi energy crosses a LL. Therefore, the field positions of the $\sigma_{xx}$ oscillation maxima are determined by the points of intersection between $\epsilon_F$ and the energy of the $i^{th}$ Landau level, given by $\epsilon_i=(i + \frac{1}{2} \pm \frac{M}{2})\hbar \omega_c$. Thus, one has to solve Eq. \ref{eq:SdH3} numerically to calculate the Fermi energy $\epsilon_F=\epsilon_F(n_{SdH},B)$. Since both the carrier concentration and the Fermi energy are unknown, we first assume a constant carrier concentration $n_{SdH}$, independent of $B$, and then adjust $n_{SdH}$ as well as $M$ simultaneously to match the measured SdH maxima of $\sigma_{xx}$ with the intersections of LLs and the Fermi energy. As pointed out recently in a similar study on Bi$_2$Se$_3$, one has to be careful when using the oscillations of other quantities than $\sigma_{xx}$ to compare with the calculated Landau spectrum \cite{Fauque:2013}. In the case of $\rho_{yx} >> \rho_{xx}$, the maxima of $\rho_{xx}$ and $\sigma_{xx}$ appear at the same fields whereas for comparable $\rho_{xx}$ and $\rho_{yx}$, the maxima of $\rho_{xx}$ and $\sigma_{xx}$ do not coincide. Since in our case $\rho_{yx} >> \rho_{xx}$ except for samples irradiated with 86 and 130 mC/cm$^2$, for which the number of observed maxima is not sufficient to perform this kind of analysis, we used $\rho_{xx}$ to perform the analysis described above. For the p-type samples, we used the $m_c$ extracted from the temperature dependence of $\Delta\rho_{xx}$ for the calculation of $\epsilon_F$. Since a reliable determination of $m_c$ for the two n-type samples ($Q=230$ and 370 mC/cm$^2$, respectively) was not possible (see above), we used $m_c=0.065$ based on the value found by K\"ohler \cite{Koehler:1976a} for comparable oscillation frequencies $\Delta(1/B)$. The longitudinal masses $m_z$ were calculated using the effective mass tensors and their dependence on $\epsilon_F$ found by K\"ohler \cite{Koehler:1976,Koehler:1976a} and are given in Tab. \ref{table1} (see the Appendix for more details).\\
Fig. \ref{fig:figure5} shows the Fermi energy obtained using Eq. \ref{eq:SdH3}, together with the Landau levels and the measured $\Delta\rho_{xx}$ for the virgin and the irradiated samples for which this analysis was possible. It turns out that the positions of all maxima can be reproduced with good accuracy for all samples. In our model, the double-peak structure observed for the p-type samples thus originates from \textit{Zeeman-splitting}. For the p-type samples (see Fig. \ref{fig:figure5} (a) to (d)), $M=1\pm0.33$ is found, i.e., either $M=0.67$ or 1.33, which agrees well with the value of $M=0.61$ found by K\"ohler for the valence band edge \cite{Koehler:1976c}. In order to make the distinction between $M=0.67$ or 1.33, measurements at higher magnetic fields are necessary to attain the quantum limit and to observe the lowest LL with index $i=0^-$. It should be noted that due to the spin-splitting factor being larger than 0.5, the peaks at higher $1/B$ values that appear not to be split are in fact a superposition of the spin-up and spin-down level of two adjacent LLs. However, due to the small amplitude $\Delta\rho_{xx}$, they appear as one single peak. Both n-type spectra could be well described assuming a spin-splitting of $M=1\pm0.43$, i.e., either $M=0.57$ or 1.43, which again agrees with the value of $M=0.53$ found by K\"ohler for the lowest conduction band \cite{Koehler:1976b}. However, since only the lower spin-split LLs are visible, the determination of $M$ and $n_{SdH}$ is less acccurate than in the case of the p-type samples. Comparison of the possible values for the $g$-factor of Bi$_2$Te$_3$, calculated from the spin-splitting factor $M$ (see Tab. \ref{table1}), shows that it lies in the same order of magnitude as the $g$-factor in Bi$_2$Se$_3$, determined by Fauqu\'e \textit{et al.} to $g=14.3$ or 28.6, respectively \cite{Fauque:2013}.\\
Assuming a 6-valley model for both the highest valence and the lowest conduction band, one has to multiply the obtained densities $n_{SdH}$ with a factor 6 in order to obtain the total carrier density. Comparison of the values for $6n_{SdH}$ with $n_H$ or $n_{H}^{\infty}$ obtained from the Hall effect shows good agreement (see Tab. \ref{table1} and Fig. \ref{fig:figure2}). This, together with the good agreement of the simulated LL spectra with the measured SdH oscillations suggests, that the electric transport in our p- as well n-type Bi$_2$Te$_3$ can be explained in terms of one single valence and conduction band, respectively, and the presence of Zeeman splitting. However, for the samples close to optimal compensation (Q=86 and 130 C/cm$^2$), two different types of carriers contribute to the conductivity.\\

\section{\label{conclusion}Conclusion}
Recently, Bi$_2$Te$_3$ has attracted enormous attention as a topological insulator (TI). However, regarding future applications of TIs, samples with an insulating bulk state are needed. We showed that it is possible to dope p-type Bi$_2$Te$_3$ in very controlled manner using electron-irradiation. The irradiation-induced defects partially compensate the Bi$_\textnormal{Te}$ antisite defects initially present in the material and convert the conduction from p- to n-type. Analysis of the Hall effect and Shubnikov-de Haas oscillations show that the electrical transport in p- (hole densities $<$ 4 x 10$^{18}$ cm$^{-3}$) as well n-type (electron densities $<$ 2.5 x 10$^{18}$ cm$^{-3}$) Bi$_2$Te$_3$ can be understood in terms of one single valence and conduction band, respectively, and the presence of Zeeman splitting. This study solely concerns the doping of bulk Bi$_2$Te$_3$; none of the irradiated samples displayed any features associated with surface carriers. However, since doping of Bi$_2$Te$_3$ using electron irradiation seems possible, it would be extremely interesting to perform these kind of irradiations on thinner Bi$_2$Te$_3$ samples on which the surface states already have been identified before irradiation. Besides doping effects, irradiation of topological insulators could give new insight on the behaviour of the surface states with respect to disorder which up to now has only been studied theoretically \cite{Schubert:2012,Guo:2010}.\\
Under high-energy electron irradiation, lattice atoms can be displaced by electrons or secondary knock-on atoms, if the energy of the projectile exceeds the threshold displacement energy $E_d$ of the different lattice atoms. Due to the high energy of the electrons of 2.5 MeV in the present study, the displacement of all different lattice atoms, i.e., Te1, Te2, Bi, as well as the displacement of the atoms on antisites is expected. However, regarding the higher total number of Te sites (ratio 3:2) and the lower displacement threshold of Te1 atoms \cite{Chaudhari:1966}, one can assume a larger number of displaced Te than Bi atoms. Since the irradiation temperature is relatively high, the displaced atoms as well as the created vacancies might form more stable interstitial or vacancy clusters or recombine with vacancies resulting in annihilation of defects or the creation of antisite defects. Both Hall effect measurements and the analysis of the observed SdH oscillations clearly indicate the electron donor character of the irradiation-induced defects. Therefore, we assume Te vacancy clusters and Te$_\textnormal{Bi}$ antisite defects, both acting as electron dopants \cite{Pecheur:1994,Hashibon:2011}, to be the predominant defects. Although the doping effect of Te and Bi interstitials has been less studied than the latter defects, Te interstitials have also been assumed to act as electron donors in a previous study on proton-irradiated Bi$_2$Te$_3$ \cite{Chaudhari:1966}. Transmission electron microscopy studies on these proton-irradiated samples showed the presence of interstitial and vacancy cluster \cite{Chaudhari:1967}. However, further detailed studies of the defects created by room and also low temperature irradiation are necessary to fully understand the damage formation in Bi$_2$Te$_3$ under irradiation.

\begin{acknowledgments}
\textbf{Acknowledgments:}\
We would like to thank A. Hruban, A. Wolos and M. Kaminska for supplying the samples and the \textit{SIRIUS} team for technical support. This work was partly supported through a SESAME grant from Region Ile-de-France.
\end{acknowledgments}

\appendix

\section{\label{AppendixA}Appendix: Calculation of carrier density from Shubnikov-de Haas oscillations}
The energy of electron states in the presence of a magnetic field $\textbf{B}=(0,0,B)$ along the $z$-direction is given by:
\begin{equation}
\epsilon(i,k_z)=(i + \frac{1}{2} \pm \frac{M}{2})\hbar\omega_{c} + \frac{\hbar^{2}}{2m_z}k_{z}^{2}
\label{eq:App1}
\end{equation}
with the cyclotron frequency $\omega_c = \frac{eB}{m_c}$, the cyclotron mass $m_c$, the longitudinal mass $m_z$, the spin-splitting factor $M=g\frac{m_c}{2 m_0}$ and the g-factor $g$. The cyclotron mass $m_c$ can be described using the effective mass tensor \textbf{m} or the inverse effective mass tensor $\boldsymbol{\alpha}=\textbf{m}^{-1}$ as $m_{c}=\sqrt{det\textbf{m}/m_{z}}$ with $m_z=\textbf{b} \cdot \textbf{m} \cdot \textbf{b}$, where $\textbf{b}$ is the unit vector of \textbf{B}.
In order to count the electronic states we assume a box in $\textbf{k}$-space of volume $V$ with sides $L_x$, $L_y$ and $L_z$. Each energy-level fixed by $(i,k_z)$ is p-fold degenerate with $p=\frac{m_c\omega_c}{2\pi}L_xL_y$ \cite{Ziman:1972}. The total number of carriers N is the sum of the carriers on each Landau level $i$
\begin{equation}
N=\sum_{i=0}^{i_{max}} \sum_{\pm M} p N_{i,k_z}^{\pm M}
\label{eq:App2}
\end{equation}
with $N_{i,k_z}^{\pm M}$ being the number of spin-up ($+M$) and spin-down ($-M$) carriers, respectively, on the $i^{th}$ Landau level. The index $i_{max}$ of the highest occupied LL is given by $i_{max}=\textnormal{floor}(\frac{\epsilon_F}{\hbar\omega_c}-\frac{1}{2}\pm\frac{M}{2})$. The wave number $k_z$ is quantized in units of $2\pi/L_z$ and varies for the $i^{th}$ Landau level between \mbox{$-k_{z,max}^{i} < k_{z} \leq k_{z,max}^{i}$} with \mbox{$k_{z,max}^{i}=k_{z,max}^{i}(\epsilon_F,\pm M)=\sqrt{\frac{2m_c}{\hbar^2}\left[\epsilon_F-(i+ \frac{1}{2} \pm \frac{M}{2})\hbar\omega_c \right]}$} calculated from Eq. \ref{eq:App1}. Therefore, for each Landau level there are
\begin{equation}
N_{i,k_z}^{\pm M}=2 \text{ } \frac{k_{z,max}^{i}(\epsilon_F,\pm M)}{(2\pi/L_z)}
\label{eq:App3}
\end{equation}
allowed states for spin-up ($+M$) and spin-down ($-M$) carriers. The total carrier concentration $n$ can then be calculated using eq. (\ref{eq:App2}) and $n=N/V$
\begin{equation}
n = \frac{1}{2\pi^2 l_{B}^{2}} \sum_{i=0}^{i_{max}}\sum_{\pm M}\sqrt{\frac{2m_z}{\hbar^2}\left[\epsilon_F-(i+ \frac{1}{2} \pm \frac{M}{2})\hbar\omega_c )\right]}
\label{eq:APP4}
\end{equation}
with $l_B=\sqrt{\hbar/eB}$.

\end{document}